# Crosslinker energy landscape effects on dynamic mechanical properties of ideal polymer hydrogels

Eesha Khare[1,2], Amadeus Alcantara[1,3,4], Nic Lee[1,5], Munir S. Skaf[4,6], Markus Buehler[1,7]*

[1]Laboratory for Atomistic and Molecular Mechanics (LAMM), Massachusetts Institute of Technology, 77 Massachusetts Avenue, Cambridge, Massachusetts, USA

[2] Department of Materials Science and Engineering, Massachusetts Institute of Technology, 77 Massachusetts Avenue, Cambridge, Massachusetts, USA

[3]Department of Computational Mechanics, School of Mechanical Engineering, University of Campinas - UNICAMP, Campinas, Sao Paulo, Brazil

[4]Center for Computing in Engineering & Sciences, CCES, University of Campinas - UNICAMP, Campinas, Sao Paulo, Brazil

[5]School of Architecture and Planning, Media Lab, Massachusetts Institute of Technology, 75 Amherst Street, Cambridge, Massachusetts, USA

[6]Institute of Chemistry, School of Mechanical Engineering, University of Campinas - UNICAMP, Campinas, Sao Paulo, Brazil

[7]Center for Computational Science and Engineering, Schwarzman College of Computing, Massachusetts Institute of Technology, 77 Massachusetts Avenue, Cambridge, Massachusetts, USA

*Corresponding author: Prof. Markus Buehler, mbuehler@mit.edu, Laboratory for Atomistic and Molecular Mechanics (LAMM), Massachusetts Institute of Technology, 77 Massachusetts Avenue, Cambridge, Massachusetts, USA

**Abstract:** Reversible crosslinkers can enable several desirable mechanical properties, such as improved toughness and self-healing, when incorporated in polymer networks for bioengineering and structural applications. In this work, we performed coarse-grained molecular dynamics to investigate the effect of the energy landscape of reversible crosslinkers on the dynamic mechanical properties of crosslinked polymer network hydrogels. We report that, for an ideal network, the energy potential of the crosslinker interaction drives the viscosity of the network, where a stronger potential results in a higher viscosity. Additional topographical analyses reveal a mechanistic understanding of the structural rearrangement of the network as it deforms and indicate that as the number of defects increases in the network, the viscosity of the network increases. As an important validation for the relationship between the energy landscape of a crosslinker chemistry and the resulting dynamic mechanical properties of a crosslinked ideal network hydrogel, this work enhances our understanding of deformation mechanisms in polymer networks that cannot easily be revealed by experiment and reveals design ideas that can lead to better performance of the polymer network at the macroscale.

## 1. Introduction

Polymer networks are important for applications ranging from biomedical engineering scaffolds,[1,2] to structural vulcanized rubber tires,[3] to active materials for gas separations.[4] This wide applicability is largely due to the structural and mechanical tailorability of polymer networks,[5] which can be tuned by a number of properties, including chain length, branching, and molecular composition. While this range of tunability enables polymer networks to be engineered for several applications, it also presents a design challenge, where predicting the mechanical properties of a polymer network from fundamental molecular design is non-trivial.[6] This prediction is even more complicated for dynamic mechanical properties of polymer networks, where important parameters such as viscoelasticity for cell engineering[7] or mechanical toughness[8] for structural engineering applications require careful accounting of time and relaxation parameters to accurately model.

One method to decouple the influence of molecular design on the dynamic mechanical properties of polymer networks involves the use of ideal reversible polymer networks, which are well-controlled polymer networks with well-defined polymer lengths, network architectures, and crosslink functionalities.[9–11] Such an ideal network allows the investigation of the effects of these aforementioned parameters on the resulting viscoelasticity of a network. For example, several experimental efforts have revealed that the viscoelasticity of polymer network hydrogels can be tuned solely by changing the crosslinker chemistry, such as metal-coordination bond[12–14] or dynamic covalent bond chemistry.[15] Several computational efforts to fundamentally understand similar relationships have also been conducted, with a few notable examples including analytical[16] or Monte-Carlo models[17] of the reaction kinetics of reversible bonds for self-healing polymer networks, theoretical models for isolating the properties of reversible crosslinks within a polymer network,[18] and hybrid computational models on the role of crosslinker strengths on the toughness of networks.[19–22] These models have enabled detailed molecular understandings on the role of crosslinker on network dynamic mechanical properties. Our previous work also sought to bridge these two approaches by using the calculated energy landscape of a few crosslinker chemistries to predict the experimental relaxation time of an ideal polymer network crosslinked by those chemistries.[23] Given these advances, in this work, we expand our previous investigation[23] to understand the impact of the energy landscape of the crosslinker on a larger crosslinked polymer network using a coarse-grained model. The coarse-grained model allows access to larger length and time scales to more directly relate the effect of the energy landscape of the individual crosslinker chemistry with the dynamic mechanical properties of the network. A thorough understanding of such a landscape can aid in the design of polymer networks with specific dynamic mechanical properties.

This work investigates the effect of the crosslinker energy landscape on the dynamic mechanical properties of an ideal reversibly crosslinked polymer network using coarse-grained simulations. Further, the molecular visualizations allow for a mechanistic understanding of the structural rearrangement of the network as it deforms as a function of the crosslinker interaction potential. Altogether, the goal of this work is to make a fundamental contribution in understanding the design principles and mechanisms of deformation that cannot easily be revealed by experiment, in order to achieve better performance of the polymer network at the macroscale and suggest future design ideas.

## 2. Results and Discussion

To model the coarse grained ideal reversible polymer network, we build a hydrogel network similar to our earlier experimental work,[13] where a tetrahedral polymer network with monodisperse 10 kDa 4-arm star-PEG molecules is constructed (**Figure 1a**). The polymer beads and water molecules are coarse-grained such that each polymer bead represents one monomer of PEG, and each water bead represents four molecules of water. To describe the chemical interactions of the polymer network, we use the force field parameters discussed in Lee et. al,[24] as these parameters have been validated for the conformation and hydrodynamics of PEG in water. The terminus of each of the arms of the PEG polymer is modeled as a second bead type, which represents the interaction between the metal ion and coordinating ligand. The bead pairwise interaction parameters are tuned to represent that interaction. Note that metal ions are not explicitly modeled, but rather are effectively modeled through changing the interaction type of the terminus (coordinating ligand) of the PEG star polymers.

No explicit bonded interactions are defined between the crosslinker bead types to allow the dynamic breaking and reforming of crosslinks. As such, to ensure that the network stably equilibrates while remaining percolated, an artificially high potential of 10000 kcal mol$^{-1}$ between the crosslinker beads is used, before being switched to the potentials explored in this study after equilibration. Note that because the crosslinker interaction potentials are not explicitly defined as being bonded, there are some network defects that emerge even in our ideal network, which are representative of defects in experimental hydrogels.[25] These defects, a dangling or clustered bond, are illustrated in **Figure 1b**. These crosslink types are considered defects because the 4-arm ideal network hydrogel is modeled such that each network junction has only two crosslinker beads in one binding interaction. If the bond is broken (i.e. there is only one bead present), the bond is considered dangling. If the bond has more than two binding partners, the bond is considered a defect with the clustered bond. In other simulations not explored in this work, the clustered defects could be considered as alternative binding arrangements for the crosslinker, such as when the metal is bound to 3 or 4 ligands. The number of defects in each equilibrated network changes as a function of the crosslinker interaction strength (**Figure 1c**). Specifically, the number of defects increases as the pair potential strength decreases, primarily driven by the increase in dangling bonds as the lower pair potential strengths. Despite the defects, all networks remain fully percolated, and the defects reported for the non-ideal network are slightly higher than other quantifications of defects in literature.[26,27]

Once the network is equilibrated, the pairwise potential of the crosslinker beads is changed to smaller interaction potentials to determine the effect of the crosslinker interaction potential (**Figure 2a**) on the resulting viscoelastic properties of the network. Due to the simulation being a shear, rather than oscillatory shear simulation, viscosity is calculated (see Methods) instead of dynamic modulus or relaxation time, and viscosity is used as a proxy for the dynamic mechanical properties of the network. The ideal and non-ideal networks were simulated under varying shear strain rates and the resulting steady state viscosities of the networks were evaluated. **Figure 2b,c** show the effect of crosslinker interaction potential on the resulting viscosity of the ideal polymer network. Both types of networks demonstrate a decrease in viscosity as the shear rate increases. This is consistent with the behavior of a non-Newtonian shear-thinning hydrogels, where viscosity decreases due to the reversible crosslinking mechanisms.[28]

In both networks, it is found that increasing the interaction potential of the crosslinker increases the viscosity of the network (**Figure 2c**). This difference in viscosity between the different interaction potentials increases as the shear rate decreases. This is expected, as the lower shear rates more directly probe a regime where the hydrogel dynamic properties should be dictated by the breaking and reforming of the crosslinker. When the energy barrier between the crosslinker is higher, it takes more energy for the bond to break, resulting in a higher viscosity of the material.

As the network deforms, the defects in the network increase (**Figure 3a**). As in **Figure 1c**, the number of defects increase more quickly for simulations with weaker interaction potentials. We computed defects based on both individual crosslinks and bond cluster, but no relevant difference was noticed.

The non-ideal polymer network has a higher viscosity than the ideal network (**Figure 3b**). In the non-ideal network, the presence of defects and multifunctional sites seem to have a larger effect on the resulting viscosity of the network, which in turn makes it difficult to directly parse out the contributions of the crosslinker potential to the network dynamics. Such network defects may be why it is more difficult to predict the imidazole network dynamics from the crosslinker chemistry energy landscape directly.[23]

In order to better understand the percolation of the polymer network as a function of crosslinker potential, topological analyses were also conducted. Modeling the network as a series of line-based graphs, percolation was analyzed by characterizing the connectivity and convolutedness of the network (**Figure 4**). Connectivity indicates the fraction of shortest-walks that are fully percolated through a network, compared to the total number of tested paths. Convolutedness is a measure of how much of the network must be explored to travel from one end to the other. Convolutedness and connectivity were found to be inversely related (**Figure 5**), indicating that networks with less connectivity require more complicated paths to traverse. No strong trend in connectivity or convolutedness was observed over time in any network. Networks increased in connectivity and decreased in average convolutedness with stronger interaction potentials.

Additional methods of topology analysis including calculation of mean curvature, atom density, continuity and valence were applied to the network (**Supplementary Information**).

## 3. Conclusions

This work set out to understand the relationship between the crosslinker chemistry energy landscape and the dynamic mechanical properties of a polymer network crosslinked by these chemistries. We found that for an ideal network, the energy potential of the crosslinker interaction has a strong effect on the resulting viscosity of the network, where a stronger interaction potential results in a higher viscosity. These viscosity values are differentiated even further as the shear rate decreases. Further, as the number of defects increases in the network, the viscosity of the network increases.

The simulations presented here offer a first step for an initial validation for the relationship between the energy landscape of a crosslinker chemistry and the resulting dynamic mechanical properties of crosslinked ideal network hydrogel. Due to the long simulation runtimes of the shear simulations, only a select number of shear-rates were tested with standard static shear simulations. Running simulations at lower speeds will take significant computational time, but may start to show a plateau in viscosity to yield a zero-shear viscosity value. The zero-shear viscosity value can more directly be compared to the relaxation time τ measured in the experimental hydrogel networks as a measure of network dynamic mechanical properties.[12,13] An alternative way to compute a more directly relatable quantity to experimental hydrogel networks would be to use oscillatory shear simulations. The resulting storage and loss modulus, and correspondingly relaxation time τ, can be more directly measured through such a simulation. This method was not explored in this present work due to challenges with long simulation times required for appropriate results. The potentials explored in this thesis are simple Lennard-Jones type interactions, and additional simulations can be conducted to evaluate the effect of changing the shape of the potential, such as by adding additional metastable states. Altogether, this section presents preliminary insights on the role the crosslinker potential plays on the dynamic viscosity of the hydrogel network and suggests several future directions for study.

## 4. Methods

Development of initial single 4-arm polymer
The ideal network polymer hydrogel was created by tessellating 128 4 arm PEG polymers. The 4-arm polymer was created via a MATLAB script, (cg_tetrahedron_v2.m which yields *singlepoly.data*), which creates the initial tetrahedral polymer structure based on the molecular weight of the 4-arm PEG polymer used in Khare et al.[23] The ends of each 4-arm polymer are described by a second type of bead, which is given a different Lennard Jones pair-wise potential. The polymer itself, excluding the crosslinker beads, is composed of all the same beads, where each polymer bead represents C-O-O, and 233 atoms comprise of one polymer with a molecular weight of 10,000 g mol$^{-1}$. The beads are described by the potential in Lee et al. who developed force field parameters to model the hydrodynamic properties of PEG in water.[24] All simulations are implemented in LAMMPS, where the harmonic bond style, harmonic angle, Fourier dihedral, and Lennard Jones/cut pair styles are used to implement the force field parameters of Lee et al.[24]l The initial polymer structure is briefly minimized and equilibrated for 2 ps with a 1 fs timestep under NVE conditions (*shortrelax.in*) to relax the polymer.

Development of polymer network
To create a percolated polymer network, this relaxed polymer was tessellated based on the crystal structure of an Ag$_2$O tetrahedral crystal (https://wiki.aalto.fi/display/SSC/Ag2O), which allows a network functionality of 2, required to mimic the functionality of the experimental polymer hydrogel. The polymer is then copied, displaced, and rotated to build the polymer lattice (*displace.in*, *combine1.in*, *displace_unit.in*, *combine2.in* sequentially). 88,613 water molecules are added (see density calculation in *fraction_constituents_peg_water.m, create_water.in*).[29] The water beads represent 4 water molecules per bead and are also described in Lee et al.[24] Then the water, and polymers are combined (*displace_linker.in*, *combine3.in* sequentially) to create the final polymer lattice with water molecules (*polybox_wlink.data*). The periodic boundary dimensions need to be adjusted by measuring the minimum and maximum dimensions of the box using VMD (VMD command: *set sel [atomselect top all] // measure minmax $sel*).

Equilibration of polymer network
The polymer network is equilibrated under fully periodic conditions with a 1 fs timestep at 296 K for 0.15 ns under NVT conditions, followed by ~10 ns equilibration at NPT until the pressure at 1 atm, total energy, and radius of gyration of a selection of polymers reaches a stable value. To keep the network fully crosslinked without defining explicit bonds between the polymers, a high interaction potential of a Lennard Jones with an energy of 10,000 kcal mol$^{-1}$, sigma of 4 Å, and cutoff of 6 Å is used. The cutoff is selected to minimize the number of cluster defects. The coordinates of the equilibrated network are saved. These equilibrated coordinates undergo further equilibration at NVT for ~1 ns at the desired interaction potential of crosslinks before undergoing shear simulation. The interaction potentials of 0, 10, 100, 500 kJ mol-1 are used with a σ distance of 4 Å and a Lennard Jones cutoff of 6 Å.

Non-equilibrium molecular dynamics shear simulations
For the non-equilibrium molecular dynamics shear simulations, varying shear rates are imposed on the equilibrated network under different crosslinker potentials (*nemd1.in*). The NVT integration with the SLLOD equations of motion (fix nvt/sllod) are used with the fix deform command to generate a velocity gradient with the desired strain rate (fix deform xy ${xyrate} erate remap v). The shear is applied in the x-y plane to represent shear stress. After a brief run of 0.5 ns after equilibration, the simulation is run for each strain rate until the viscosity plateaus for at least 50 ns, resulting in shear simulations running from 100 ns to over 1500 ns. The viscosity is calculated by dividing the xy component of the pressure tensor by the strain rate (v_srate) and the length of the box ly (-pxy/v_srate/ly). A timestep of 2 fs is used across all simulations, and viscosity values are averaged every 0.2 ns. Defects, specified as crosslinks without exactly 1 neighboring crosslink within 5 Å, are calculated every 200 ns using a MATLAB script.

Topological analysis
To investigate if and how the different interatomic potentials influence the geometry and topology of the devised models during shear deformation, we performed topological analyses. For this, the polymer networks were imported as graphs and converted to line-based geometries. Each line consisted of a straight edge between two nodes representing the position of atoms in the polymer network. All topological analyses were conducted using the Visual Effects software Houdini (SideFX, Toronto).

To visualize the networks, colors were assigned to each node and edge according to their distance from the origin in cartesian space. Edges were converted to 3D mesh geometries using a marching cubes algorithm[30] with a radius of 0.5 Å for general visualizations.

In order to assess percolation within the polymer network, shortest walks were computed between distant points at the network's periphery (**Figure 4**). The successful calculation of a path between these two points indicates network percolation. A connectivity matrix was constructed from 1020 point-to-point connections longer than 100 Å along the network's periphery, and the endpoints of each line were used as start and end points for the calculation of a shortest walk using the heat equation for calculating geodesic distance.[31] For every successful path, the path's length was divided by the distance between the end points in order to create a metric of "convolutedness". Higher measures of network convolutedness indicate that to travel from one end of the network to the other, a greater portion of the network must be explored. This occurs when there are fewer viable paths through the network.

Connectivity was measured as a fraction of viable shortest-walks to the total number of tested paths in the original connectivity matrix (**Figure 4**). A connectivity value of 1 indicates that every one of the 1020 pairs of start and end points tested resulted in a viable path through the network, while lower values indicate less connectivity and a greater number of isolated regions in the network.

**Funding and Acknowledgments**

E. K. acknowledges the NSF Graduate Research Fellowship Program and MIT Office of Graduate Education. The authors acknowledge financial support from NIH (U01EB014976, 1R01AR07779, 1R01AR077793, and R01HL159094), ARO (W911NF1920098), the NSF (CMMI 1548571 and DMR 2105150), ONR (N00014-19-1-2375 and N00014-20-1-2189), the São Paulo Research Foundation (FAPESP) grants #2018/18503-2, #2022/03410-4, and the Center for Computing in Engineering & Sciences (CCES/UNICAMP) FAPESP grant #2013/08293-7.

**References**

1   T. R. Hoare and D. S. Kohane, *Polymer (Guildf).*, 2008, **49**, 1993–2007.
2   R. Langer and D. A. Tirrell, *Nature*, 2004, **428**, 487–492.
3   P. J. Flory, *Chem. Rev.*, 1944, **35**, 51–75.
4   S. Kim and Y. M. Lee, *Prog. Polym. Sci.*, 2015, **43**, 1–32.
5   M. Rubinstein and S. Panyukov, *Macromolecules*, 2002, **35**, 6670–6686.
6   C. P. Broedersz and F. C. Mackintosh, *Rev. Mod. Phys.*, 2014, **86**, 995–1036.
7   B. K. Chan, C. C. Wippich, C. J. Wu, P. M. Sivasankar and G. Schmidt, *Macromol. Biosci.*, 2012, **12**, 1490–1501.
8   X. Zhao, *Proc. Natl. Acad. Sci. U. S. A.*, 2017, **114**, 8138–8140.
9   G. A. Parada and X. Zhao, *Soft Matter*, 2018, **14**, 5186–5196.
10  F. Tanaka and S. F. Edwards, *J. Nonnewton. Fluid Mech.*, 1992, **43**, 289–309.
11  F. Tanaka and S. F. Edwards, *J. Nonnewton. Fluid Mech.*, 1992, **43**, 247–271.
12  D. E. Fullenkamp, L. He, D. G. Barrett, W. R. Burghardt and P. B. Messersmith, *Macromolecules*, 2013, **46**, 1167–1174.
13  S. C. Grindy, R. Learsch, D. Mozhdehi, J. Cheng, D. G. Barrett, Z. Guan, P. B. Messersmith and N. Holten-Andersen, *Nat. Mater.*, 2015, **14**, 1210–1216.
14  T. Rossow and S. Seiffert, *Polym. Chem.*, 2014, **5**, 3018–3029.
15  B. Marco-Dufort, R. Iten and M. W. Tibbitt, *J. Am. Chem. Soc.*, 2020, **142**, 15371–15385.
16  E. B. Stukalin, L. H. Cai, N. A. Kumar, L. Leibler and M. Rubinstein, *Macromolecules*, 2013, **46**, 7525–7541.
17  D. Amin, A. E. Likhtman and Z. Wang, *Macromolecules*, 2016, **49**, 7510–7524.
18  X. Zhang, Y. Vidavsky, S. Aharonovich, S. J. Yang, M. R. Buche, C. E. Diesendruck and M. N. Silberstein, *Soft Matter*, 2020, **16**, 8591–8601.
19  B. V. S. Iyer, V. V. Yashin, M. J. Hamer, T. Kowalewski, K. Matyjaszewski and A. C. Balazs, *Prog. Polym. Sci.*, 2015, **40**, 121–137.
20  A. C. Balazs, *Mater. Today*, 2007, 10, 18–23.
21  G. V Kolmakov, K. Matyjaszewski and A. C. Balazs, *ACS Nano*, 2009, **3**, 885–892.
22  Y. Liu, J. Aizenberg and A. C. Balazs, *Nanomaterials*, 2021, **11**, 2764.
23  E. Khare, S. A. Cazzell, J. Song, N. Holten-Andersen and M. J. Buehler, *Proc. Natl. Acad. Sci. U. S. A.*, 2023, **120**, e2213160120.
24  H. Lee, A. H. De Vries, S. J. Marrink and R. W. Pastor, *J. Phys. Chem. B*, 2009, **113**, 13186–13194.
25  F. Di Lorenzo and S. Seiffert, *Polym. Chem.*, 2015, 6, 5515–5528.
26  M. Zhong, R. Wang, K. Kawamoto, B. D. Olsen and J. A. Johnson, *Science (80-. ).*, 2016, **353**, 1264–1268.
27  F. Lange, K. Schwenke, M. Kurakazu, Y. Akagi, U. Il Chung, M. Lang, J. U. Sommer, T. Sakai and K. Saalwächter, *Macromolecules*, 2011, **44**, 9666–9674.
28  M. H. Chen, L. L. Wang, J. J. Chung, Y. H. Kim, P. Atluri and J. A. Burdick, *ACS Biomater. Sci. Eng.*, 2017, **3**, 3146–3160.
29  A. C. S. Alcântara, L. C. Felix, D. S. Galvão, P. Sollero and M. S. Skaf, *Materials (Basel).*, 2022, **15**, 2274.
30  T. S. Newman and H. Yi, *Comput. Graph.*, 2006, **30**, 854–879.
31  K. Crane, C. Weischedel and M. Wardetzky, *ACM Trans. Graph.*, 2013, **32**, 1–11.

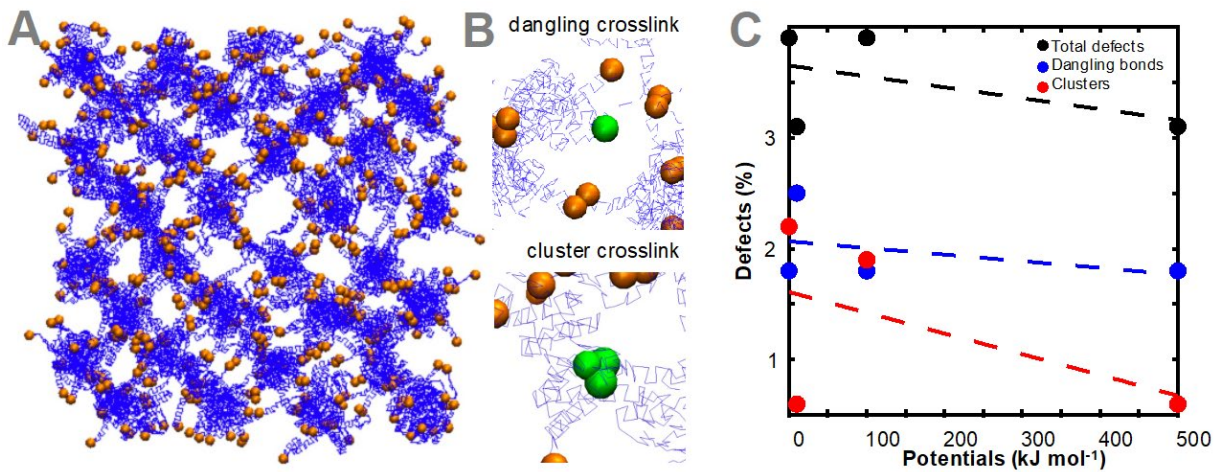

**Figure 1. Polymer network setup.** A) Equilibrated polymer network with polymer (blue) and crosslinker (orange) demonstrated. Crosslinker beads are enlarged and water beads are not shown for clarity. Polymer network is based on the crystal unit cell of an $Ag_2O$ tetrahedral crystal. B) Example of defects (green beads) in network include the dangling crosslinks (no bonding partner) or clusters of bonds (more than 1 bonding partner). The ideal network has 2.7% defects based on the individual bond crosslinks. C) Percent of defects in network based on strength of interaction potential between crosslinking beads. Increasing the strength of the interaction potential reduces the total number of defects by reducing the number of dangling bonds present in the network.

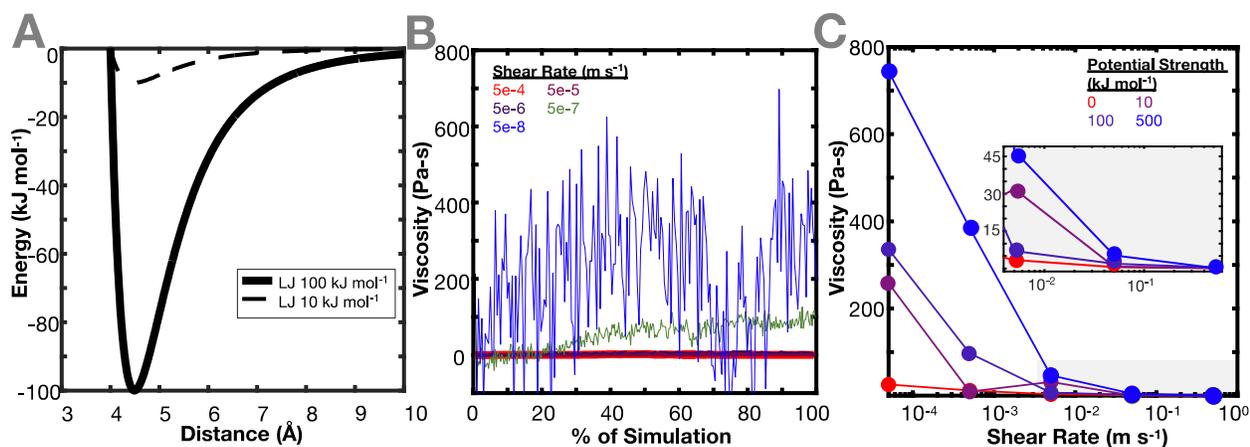

**Figure 2. Effect of crosslinker energy landscape on viscosity of polymer network.** A) Examples of changing crosslinker energy landscape potentials applied in this work. A standard Lennard-Jones interaction is used, with changing interaction strengths, which changes both the depth of the well, and its width. B) Representative data collection of viscosity over simulation time, normalized by the total length of the simulation at each respective shearing speed. The data shown is for the Lennard-Jones crosslinker interaction strength of 100 kJ mol$^{-1}$. Once the final 50 ns of the viscosity plateaus, the values are averaged and plotted in C), which shows the effect of the crosslinker interaction potential on the resulting viscosity of the network as a function of shear rate.

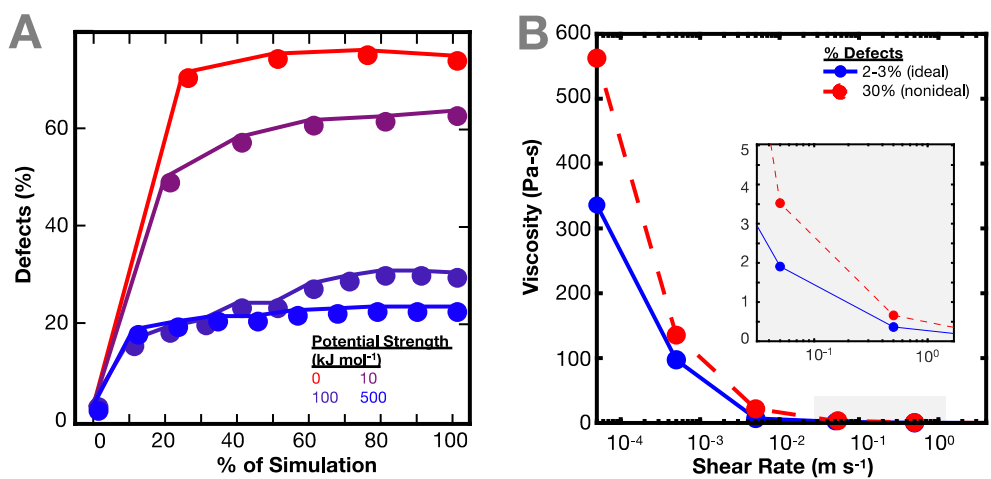

**Figure 3. Effect of defects on viscosity of polymer network.** A) Defects in network based on individual crosslinks for $10^{-9}$ fs$^{-1}$ engineering strain rate across different interaction potentials. The defects increase over simulation time. B) Comparison of viscosity of network with 2-3 % initial defects versus ~ 30% initial defects for an interaction potential of 100 kJ mol$^{-1}$.

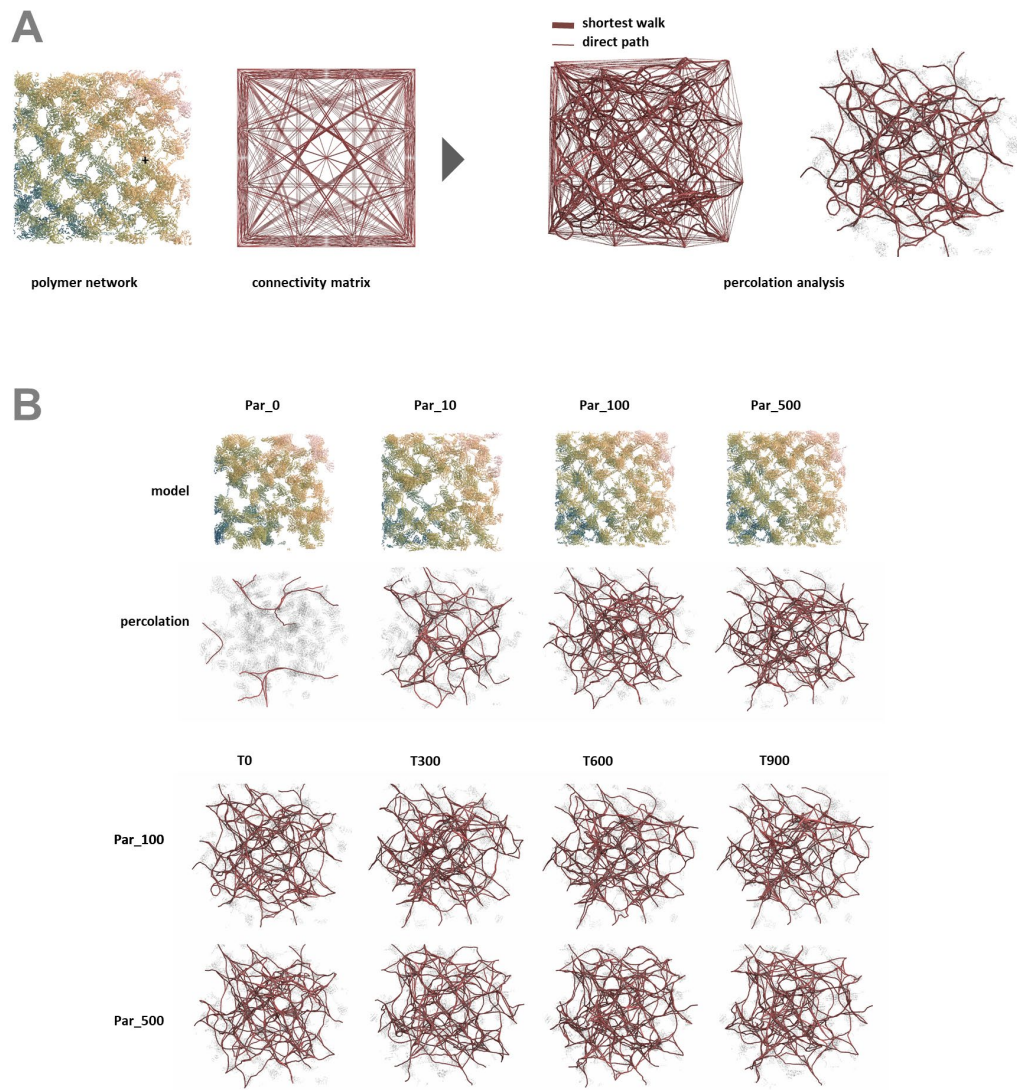

**Figure 4. Percolation was assessed within each polymer network at multiple time steps under deformation.** A) Percolation was measured by comparing the number of possible paths between distant points on the network's periphery to the number of viable shortest walks through the network between those points. B) Each network demonstrated different degrees of connectivity, and the number and form of viable paths through the network changed under deformation.

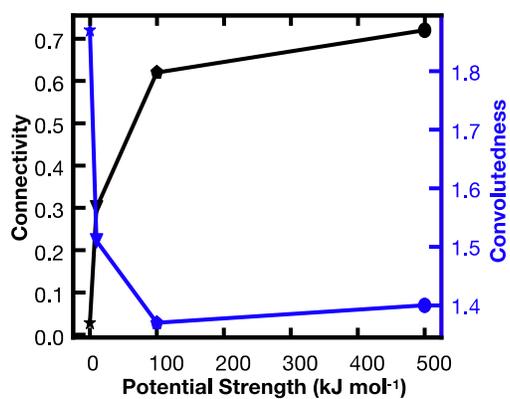

**Figure 5. Connectivity vs. Convolutedness for models with interaction potentials of 0, 10, 100, 500 kJ mol$^{-1}$.** As the interaction potential increases, the connectivity of the network goes up, while the convolutedness goes down. This implies that a stronger potential holds the network together more robustly.

# Crosslinker energy landscape effects on dynamic mechanical properties of ideal polymer hydrogels

Eesha Khare[1,2], Amadeus Alcantara[1,3,4], Nic Lee[1,5], Munir S. Skaf[4,6], Markus Buehler[1,7]*

[1]Laboratory for Atomistic and Molecular Mechanics (LAMM), Massachusetts Institute of Technology, 77 Massachusetts Avenue, Cambridge, Massachusetts, USA

[2] Department of Materials Science and Engineering, Massachusetts Institute of Technology, 77 Massachusetts Avenue, Cambridge, Massachusetts, USA

[3]Department of Computational Mechanics, School of Mechanical Engineering, University of Campinas - UNICAMP, Campinas, Sao Paulo, Brazil

[4]Center for Computing in Engineering & Sciences, CCES, University of Campinas - UNICAMP, Campinas, Sao Paulo, Brazil

[5]School of Architecture and Planning, Media Lab, Massachusetts Institute of Technology, 75 Amherst Street, Cambridge, Massachusetts, USA

[6]Institute of Chemistry, School of Mechanical Engineering, University of Campinas - UNICAMP, Campinas, Sao Paulo, Brazil

[7]Center for Computational Science and Engineering, Schwarzman College of Computing, Massachusetts Institute of Technology, 77 Massachusetts Avenue, Cambridge, Massachusetts, USA

*Corresponding author: Prof. Markus Buehler, mbuehler@mit.edu, Laboratory for Atomistic and Molecular Mechanics (LAMM), Massachusetts Institute of Technology, 77 Massachusetts Avenue, Cambridge, Massachusetts, USA

# Supporting Information

**Topology Measurements**

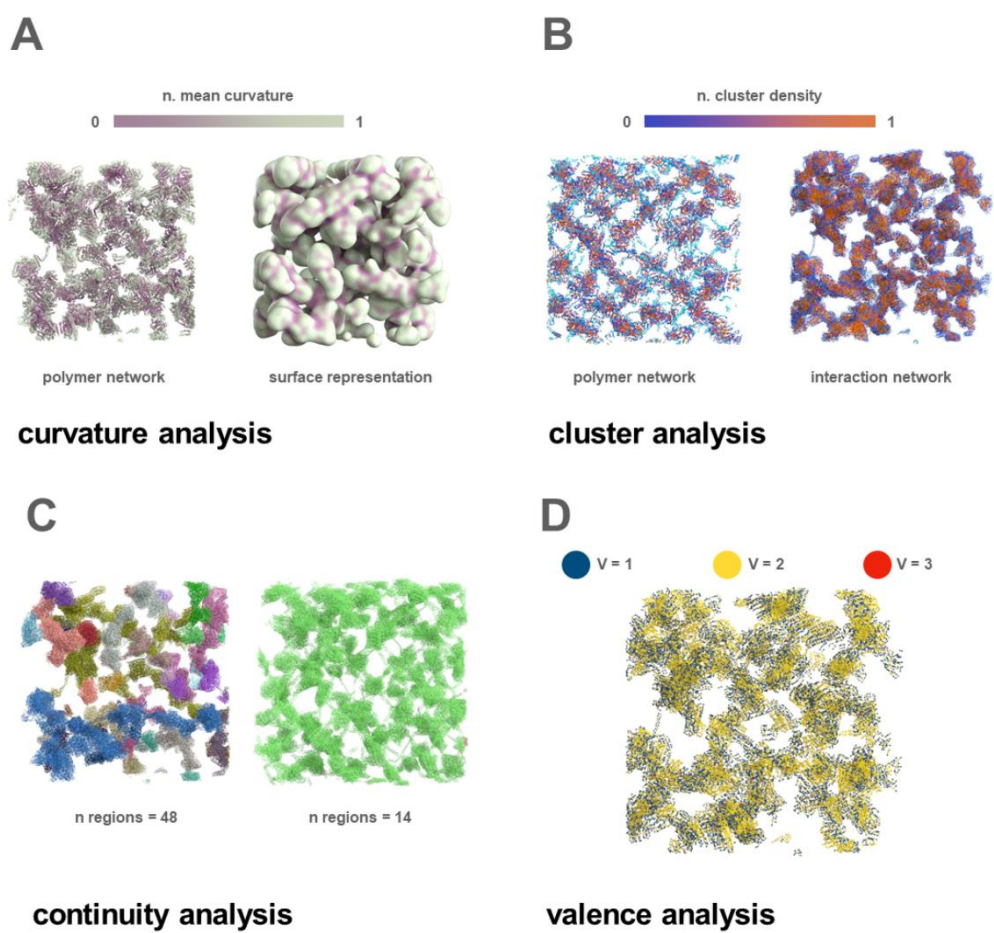

**Figure S1.** Additional methods of topology analysis were applied to the network. (A) The mean curvature of a surface representation of the network was computed to indicate regions of concave and convex curvature. (B) The density of atoms in the network was measured by the number of possible interactions within a 5 A radius. (C) Continuity was assessed based on the number of regions formed when all atoms within 5 A were connected with one another. (D) Valence was measured according to the number of atom-to-atom interactions within the network.

**Curvature**

Mean curvature was measured for the polymer network by creating a mesh surface representation of the network and sampling curvature at each vertex. First, a signed distance field (SDF) [1] was created from the point cloud of atom positions in the polymer network. Voxels in the SDF were assigned a radius of 1 A. This preliminary voxel field was expanded by increasing the radius engaged by each voxel by a factor of ten before applying a gaussian smoothing function for ten iterations [2]. The SDF was then converted to a quadrilateral mesh using the marching cubes algorithm, which resulted in the creation of a coarse-surface representation of the polymer network.

Mean curvature is a signed value that measures curvature in two principal directions perpendicular to the surface normal. While Gaussian curvature $K=\kappa_1\kappa_2$ is the square of the geometric mean of the principal curvatures $\kappa_1$ and $\kappa_2$, mean curvature $H=(\kappa_1+\kappa_2)/2$ is the arithmetic mean of $\kappa_1$ and $\kappa_2$. Principal curvatures describe the maximum $\kappa_1$ and minimum $\kappa_2$ curvature at every point on the surface. Low mean curvature in the mesh representation of the network occurs in regions where the network narrows into thin connections, while high mean curvature occurs in convex regions where atoms are clustered.

**Valence**
Valence is the number of direct connections for each atom in the polymer network according to the imported model. The vast majority of atoms in the network have a valence of 2, while a small number have either 1 or 3 connections.

**Density**
Density indicates the number of atoms within a region of the network. High density indicates a large number of atoms in a relatively small area, referred to as a cluster. A measurement of atom density was created by adding additional connections to all adjacent atoms within 5 A of one another and measuring the connections at each atom. The highest observed density in the network is 30 while the lowest is 1.

**Continuity**
Continuity indicates the number of connected regions in the network after connections are added to every atom within 5 A of another. A higher number of regions may indicate a more discontinuous network, or may simply indicate additional breakages in already isolated regions.

## Supplementary Citations

1.Oleynikova, H., Millane, A., Taylor, Z., Galceran, E., Nieto, J., & Siegwart, R. (2016). Signed distance fields: A natural representation for both mapping and planning. In RSS 2016 workshop: geometry and beyond-representations, physics, and scene understanding for robotics. University of Michigan.
2.Getreuer, P. (2013). A survey of Gaussian convolution algorithms. Image Processing On Line, 2013, 286-310.